\documentclass{aa}

\usepackage{graphicx}
\usepackage{txfonts}
\usepackage{natbib}
\usepackage{amsmath,amssymb}
\usepackage{hyperref}
\usepackage{bm}
\usepackage{enumitem}
\usepackage[normalem]{ulem}
\usepackage{xcolor}
\usepackage{soul}

\newcommand{\ha}{\mathrm{H\alpha}}
\newcommand{\rhokq}{\rho^K_Q}
\newcommand{\jkq}{J^K_Q}

\newcommand{\aju}{\alpha_{u}J_{u}}
\newcommand{\ajl}{\alpha_{\ell}J_{\ell}}

\newcommand{\dtrans}[2]{{#1}\rightleftharpoons{#2}}

\begin{document}

\title{Three-dimensional radiative transfer modeling of the scattering polarization in the solar H$\alpha$ line}

\titlerunning{3D modeling of H$\alpha$ scattering polarization}

\author{
Jaume Jaume Bestard\inst{1}
\and
Ji\v{r}\'i \v{S}t\v{e}p\'an\inst{2}\corrauth{jiri.stepan@asu.cas.cz}
\and
Javier Trujillo Bueno\inst{1,3,4}
}

\institute{
Instituto de Astrof\'isica de Canarias, E-38205 La Laguna, Tenerife, Spain
\and
Astronomical Institute of the Academy of Sciences, 25165 Ond\v{r}ejov, Czech Republic.
\and
Departamento de Astrof\'isica, Universidad de La Laguna, E-38206 La Laguna, Tenerife, Spain
\and
Consejo Superior de Investigaciones Cient\'ificas, Spain
}

\authorrunning{J. Jaume Bestard et al.}

\date{Received XXXX; accepted XXXX}

\abstract
{
The H$\alpha$ line is one of the most important diagnostics of the solar chromosphere, but interpreting its scattering polarization remains challenging because of the combined action of magnetic fields, macroscopic motions, and the strongly inhomogeneous 3D structure of the atmosphere.
}
{
We investigate the formation of the intensity and linear polarization of the solar H$\alpha$ line in a 3D magnetohydrodynamic model of the solar atmosphere and assess the diagnostic information encoded in the observed scattering polarization signals.
}
{
We solved the full 3D non-local thermodynamic equilibrium (non-LTE) polarized radiative transfer problem for a multilevel hydrogen atom using the PORTA code in a simulation that includes nonequilibrium hydrogen ionization. The calculations account for atomic polarization, Hanle-effect magnetic depolarization, horizontal radiative transfer, and macroscopic plasma velocities. The synthetic Stokes profiles were degraded to the spatial, spectral, and polarimetric resolution of available ZIMPOL observations to enable a direct comparison with measurements.
}
{
The calculated center-to-limb variation of the H$\alpha$ scattering polarization reproduces the observed overall trend and yields polarization amplitudes comparable to the observations. The simulations reveal a large spatial variability of the polarization signals arising from symmetry-breaking effects caused by horizontal inhomogeneities and atmospheric dynamics. The synthetic profiles naturally exhibit a variety of line-core asymmetries. By classifying the emergent $Q/I$ profiles according to their degree of asymmetry, we find that their morphology is statistically correlated with the magnetic-field strength in the H$\alpha$ formation region and is consistent with Hanle-effect depolarization. Multi-peaked profiles preferentially arise in regions with stronger magnetic fields, whereas asymmetric single-peaked profiles are associated with weaker magnetic fields exhibiting comparatively smaller vertical variations in strength.
}
{
The scattering polarization of the H$\alpha$ line is governed by the complex interplay between magnetic fields, velocity gradients, and 3D radiative transfer effects. Our results demonstrate the importance of accounting for 3D radiative transfer effects when interpreting H$\alpha$ spectropolarimetric observations and confirm the diagnostic potential of this line for probing the magnetism and dynamics of the solar chromosphere.
}


\maketitle
\nolinenumbers

\section{Introduction}\label{sec:intro}

Magnetic fields emerging from the solar photosphere expand into the chromosphere, where they play a fundamental role in shaping the thermal structure, dynamics, and energetics of plasma. Diagnosing chromospheric magnetic fields is therefore essential for understanding the physical processes that couple the lower and upper layers of the solar atmosphere. Spectropolarimetric observations provide one of the few direct diagnostics of chromospheric magnetic fields. Their interpretation, however, remains challenging because the physical mechanisms responsible for generating polarization in chromospheric spectral lines are intrinsically complex, while the chromosphere itself is highly structured, dynamic, and far from local thermodynamic equilibrium \citep[LTE; e.g., the review by][]{2022ARA&A..60..415T}.

During the last decade, substantial progress has been achieved in modeling polarized line formation in realistic 3D magnetohydrodynamic (MHD) simulations of the solar atmosphere. Such investigations have demonstrated that horizontal inhomogeneities and macroscopic velocity gradients break the axial symmetry of the radiation field, producing significant modifications of the scattering polarization signals of spectral lines \cite[e.g.,][]{2011ApJ...743...12M,2015lya,2016ApJ...826L..10S,2021ApJ...909..183J}. Time-dependent studies have further shown that the temporal evolution of the chromosphere can substantially reduce the amplitudes of the emergent polarization signals \citep{2013ApJ...764...40C,2018ApJ...863..164D}. These findings highlight the importance of accounting for 3D radiative transfer effects under non-local thermodynamic equilibrium (non-LTE) conditions when interpreting chromospheric spectropolarimetric observations.

Among chromospheric diagnostics, the hydrogen H$\alpha$ line occupies a privileged position. Owing to its large opacity and extended formation region, it is sensitive to the magnetic, dynamic, and thermodynamic properties of a significant fraction of the chromosphere. Its scattering polarization signals have been observed in quiet-Sun regions and constitute one of the conspicuous features of the second solar spectrum \citep{2000sss..book.....G,2022A&A...659A.179J}. In particular, the line-center linear polarization profile exhibits well-known asymmetry near the solar limb \citep{2011ASPC..437..117S}, whose physical origin remains a matter of considerable interest.

Recently, \citet{2022A&A...659A.179J} presented high-sensitivity full-Stokes observations of the H$\alpha$ line obtained with the Zurich Imaging Polarimeter (ZIMPOL), reaching a polarimetric sensitivity on the order of $10^{-4}$. These observations reveal substantial spatial variability in scattering polarization signals, together with asymmetric $Q/I$ profiles at both low and moderate spatial resolutions. Given that the linear polarization of H$\alpha$ is primarily generated by scattering processes, such observations suggest an important role for symmetry-breaking effects produced by spatial variations of temperature, density, velocity, and magnetic field throughout the chromosphere.

The origin of the observed line-core asymmetry (LCA) has been investigated previously using 1D non-LTE radiative transfer calculations. In particular, \citet{2010ApJ...711L.133S} and \citet{2011ASPC..437..117S} showed that strong magnetic-field gradients within the H$\alpha$ formation region can generate asymmetric $Q/I$ profiles similar to those observed near the solar limb. However, 1D models cannot account for the horizontal inhomogeneities and complex dynamics that characterize the real chromosphere. Whether the observed asymmetries originate primarily from magnetic gradients, atmospheric dynamics, or the interplay between both effects in a realistic 3D atmosphere remains an open question.

In this paper, we investigate the scattering polarization of the H$\alpha$ line by solving the full 3D non-LTE polarized radiative transfer problem in a realistic MHD model of the solar atmosphere. The calculations are carried out with the PORTA code \citep{2013A&A...557A.143S}, specifically designed for large-scale 3D polarized radiative transfer applications. We present synthetic intensity and polarization profiles, compare them with recent ZIMPOL observations, and analyze the relationship between the resulting line-core asymmetries and the physical properties of the model atmosphere. Our aim is to investigate the diagnostic potential of the H$\alpha$ line for probing the magnetic and dynamic structure of the solar chromosphere.

\section{The physical problem}

\subsection{Atomic model}
\label{sb:atom_halpha}

The H$\alpha$ line is a subordinate transition connecting the $n=2$ and $n=3$ levels of hydrogen. Since the lower level of H$\alpha$ is the upper level of Ly$\alpha$, while its upper level is shared with Ly$\beta$, a realistic treatment of the scattering polarization of H$\alpha$ requires a self-consistent description of the first three principal levels of hydrogen and of the radiative coupling among the Ly$\alpha$, Ly$\beta$, and H$\alpha$ lines.

The atomic model adopted in this work consists of the nine fine-structure levels considered by \citet[][see their Fig.~1 for the corresponding Grotrian diagram]{2011ApJ...732...80S}. The corresponding radiative transitions form the Ly$\alpha$, Ly$\beta$, and H$\alpha$ lines. As discussed by \citet{2011ApJ...732...80S}, the inclusion of the $n=4$ level has only a minor impact on the scattering polarization of H$\alpha$. Although transitions of the Paschen series could in principle affect the atomic polarization of the $n=3$ levels, their expected contribution is small compared with the substantial increase in computational complexity required by a larger atomic model. We therefore restrict our calculations to the first three principal levels.

Hyperfine structure is neglected throughout this investigation. Previous studies have shown that its impact on the polarization of hydrogen lines is negligible, both for Ly$\alpha$ \cite[][]{1982SoPh...78..157B} and for the Balmer series \cite[][]{1986A&A...156...79B}. We therefore adopt the fine-structure approximation employed by \citet{2011ApJ...732...80S}.

Transitions involving only levels with $J=1/2$ cannot carry atomic alignment and therefore do not contribute to the linear scattering polarization. The remaining transitions contribute through selective emission and absorption processes. In particular, the alignment of the $2p_{3/2}$ level gives rise to dichroism, which may significantly modify the emergent polarization signals \cite[][]{1997ApJ...482L.183T}.

The wavelength separations and relative strengths of the fine-structure components are shown in Fig.~2 of \citet{2011ApJ...732...80S}. For H$\alpha$ they are on the order of $10^{-1}$\AA, whereas for Ly$\alpha$ and Ly$\beta$ they are approximately two orders of magnitude smaller. Under solar chromospheric conditions, the individual components are strongly blended. For example, at $T=5000\,$K the Doppler width of an individual H$\alpha$ component is approximately $0.2\,$\AA, comparable to the separation between the most distant transitions.

The strength of a radiative transition $\dtrans{\ajl}{\aju}$ is characterized by
\begin{equation}
\mathcal{S}(\aju,\ajl)=
(2J_u+1)
\frac{3hc^3}{64\pi^4\nu_{u\ell}^3}
A_{u\ell},
\end{equation}
which measures the coupling of the transition to the radiation field through the electric dipole interaction.

Quantum interference between different $J$ levels is known to play an important role in the polarization of strong resonance lines such as Ly$\alpha$ and Mg\,{\sc ii} $h$ and $k$ \cite[][]{2012ApJ...755L...2B,2014A&A...564A..16B}. However, for the subordinate H$\alpha$ line this effect can be safely neglected \cite[][]{2011ApJ...732...80S}. Likewise, magnetic-field-induced coherences among the $n=3$ fine-structure levels become significant only for field strengths on the order of $100\,$G or larger. Since the average magnetic field strength in the H$\alpha$ formation region of our model is about $20\,$G (Fig.~\ref{fig:model_height}), neglecting such coherences is a reasonable approximation for the quiet-Sun conditions considered here.
Further details of the density-matrix formulation are given in Appendix~A, while the radiative-transfer formulation and the radiation-field tensors are summarized in Appendix~B.

\subsection{Atmospheric model}

The radiative-transfer calculations were carried out in the 3D MHD simulation of \citet{2016A&A...585A...4C}, computed with the Bifrost code \cite[][]{2011A&A...531A.154G}. This simulation includes nonequilibrium hydrogen ionization, which is important for obtaining realistic electron and hydrogen densities in the chromosphere \cite[][]{1980A&A....87..229K,2002ApJ...572..626C}. Following \citet{2012ApJ...749..136L}, the polarized radiative-transfer problem was solved assuming statistical equilibrium for the atomic density matrix, while retaining the nonequilibrium ionization state provided by the MHD simulation.

The H$\alpha$ line forms over a broad range of chromospheric heights. Figure~\ref{fig:formation-height} shows the distribution of line-center formation heights, defined by the condition $\tau_\mu(\lambda_0)=1$, for disk-center and near-limb lines of sight (LOSs). At disk center, the line center typically forms between 400 and 2000~km, with a secondary peak around 500~km associated with the well-known H$\alpha$ opacity gap \cite[][]{1972SoPh...22..344S}. Toward the limb, the contribution from lower layers becomes progressively less important, and the line formation shifts to the upper chromosphere.

\begin{figure}[t]
  \centering
  \includegraphics[width=0.5\textwidth]{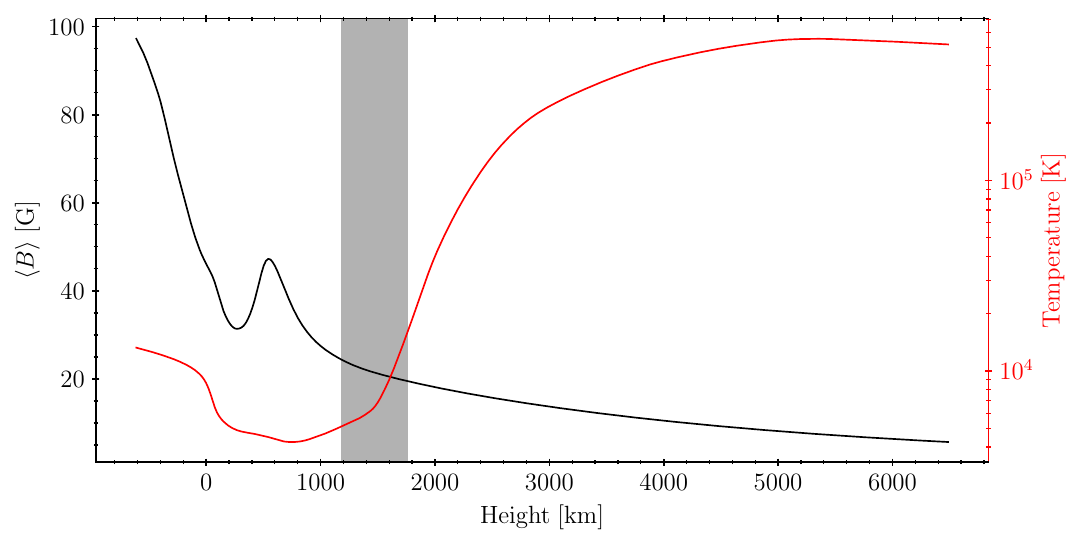}
\caption{
Height variation of the horizontally averaged magnetic field strength (black curve) and temperature (red curve) in the 3D MHD model atmosphere. The gray-shaded region indicates the range of mean $\ha$ line-center formation heights for LOS inclinations between $\mu=1$ and $\mu=0.1$.
}
  \label{fig:model_height}
\end{figure}

\begin{figure}[t]
  \centering
  \includegraphics[width=\columnwidth]{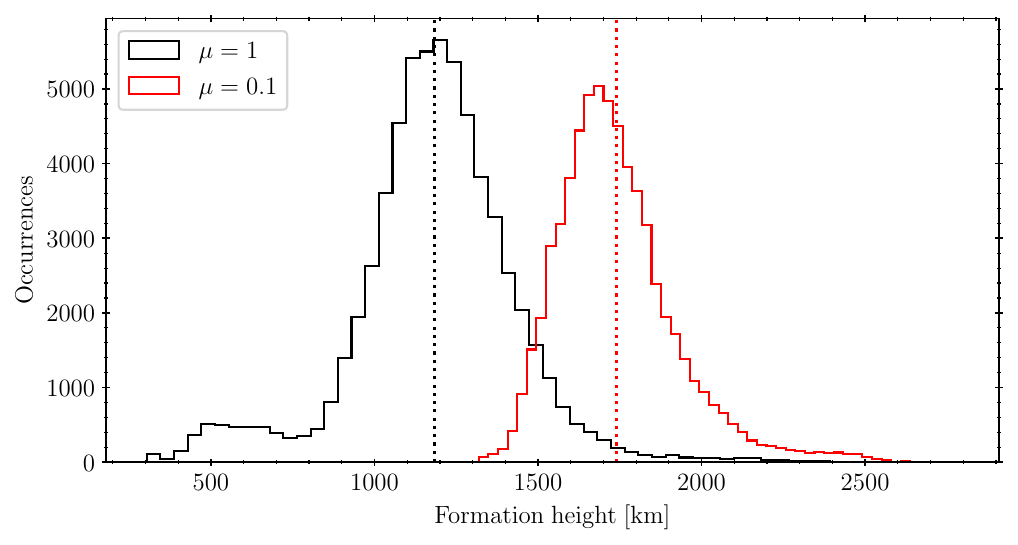}
\caption{
Histogram of the $\ha$ line-center formation heights, defined by $\tau_\mu(\lambda_0)=1$, at all $(x,y)$ locations in the 3D MHD model. The black and red histograms correspond to $\mu=1$ and $\mu=0.1$, respectively. The vertical dotted lines indicate the mean formation heights.
}
  \label{fig:formation-height}
\end{figure}

The corresponding range of formation heights is indicated in Fig.~\ref{fig:model_height}, together with the height variation of the horizontally averaged temperature and magnetic field strength. The mean magnetic field in the H$\alpha$ formation region is approximately 20 G, although localized structures in the middle chromosphere can reach several hundred gauss \cite[][]{2016ApJ...826L..10S}. Outside such regions, the atmospheric conditions are representative of the quiet Sun.

\section{The self-consistent solution\label{sec:fs}}

\begin{figure}[t]
	\centering
	\includegraphics[width=1\columnwidth]{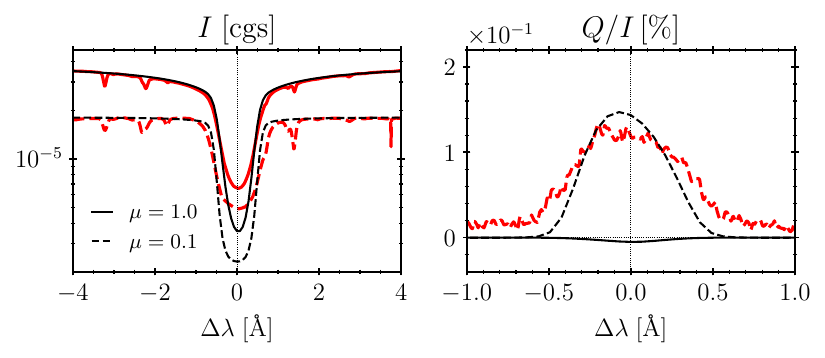}
	\caption{
		Spatially averaged Stokes $I$ and $Q/I$ profiles of the emergent radiation for the $\mu=1$ and $\mu=0.1$ LOSs. The red curves show the observed profiles reported by \cite{1984sfat.book.....K} for $\mu=1$ (solid lines) and by \cite{2000sss..book.....G} for $\mu=0.1$ (dashed lines). The black curves show the corresponding synthetic profiles obtained from the 3D MHD model. The reference direction for positive Stokes $Q$ is parallel to the nearest solar limb.
	}
	\label{fig:avg_mus_profiles}
\end{figure}

After obtaining the self-consistent solution of the statistical equilibrium equations with PORTA, we computed the emergent Stokes profiles for different lines of sight throughout the field of view. The calculations presented here employ the long-characteristics formal solver implemented in PORTA \citep{2021ApJ...912...63D}, which minimizes the numerical diffusion associated with interpolation procedures in short-characteristics methods \citep{1988JQSRT..39...67K}. Further details of the numerical setup and boundary conditions are given in Appendix~\ref{sb:boundary_halpha}. The preconditioning technique and further details of the implementation in the PORTA code are described in Appendices~\ref{sb:precond_halpha} and \ref{sb:implementation_halpha}, respectively.

Figure~\ref{fig:avg_mus_profiles} compares the spatially averaged synthetic profiles with observations at disk center ($\mu=1$) and near the limb ($\mu=0.1$). The synthesized intensity profiles are systematically narrower and deeper than the observed ones. This discrepancy is consistent with previous studies based on the same 3D MHD simulation, which indicate that the model lacks part of the small-scale dynamics and unresolved motions present in the real chromosphere \citep{2016A&A...585A...4C}. Additional differences may arise from inaccuracies in the hydrogen density stratification and from the unavoidable simplifications of the adopted atomic model.

The agreement is significantly better for the linear polarization. Near the limb, the calculated $Q/I$ amplitudes are comparable to the observed ones and show broad agreement with the overall observed profile shape. This result contrasts with previous 1D calculations \citep{2011ASPC..437..117S}, which yielded polarization amplitudes substantially larger than observed owing to the absence of Hanle depolarization and 3D symmetry-breaking effects. The present calculations therefore indicate that horizontal inhomogeneities, velocity fields, and magnetic depolarization are all important ingredients for understanding the scattering polarization of H$\alpha$.

An interesting property of both the observed and synthetic profiles is the presence of a LCA. The observed profile shown in Fig.~\ref{fig:avg_mus_profiles} exhibits a slight enhancement on the blue side with respect to the red side, while the synthetic profiles also exhibit line-core asymmetries, which
depend on the LOS and the atmospheric structures sampled. The asymmetry obtained in the 3D calculations is qualitatively similar to that predicted by \citet{2010ApJ...711L.133S} using 1D atmospheric models with strong magnetic-field gradients. The extent to which such asymmetries originate from either magnetic effects alone or their interplay with chromospheric dynamics and horizontal inhomogeneities is discussed in Sect.~\ref{sec:asym}.

\begin{figure*}[t!]
  \centering
  \includegraphics[width=0.8\textwidth]{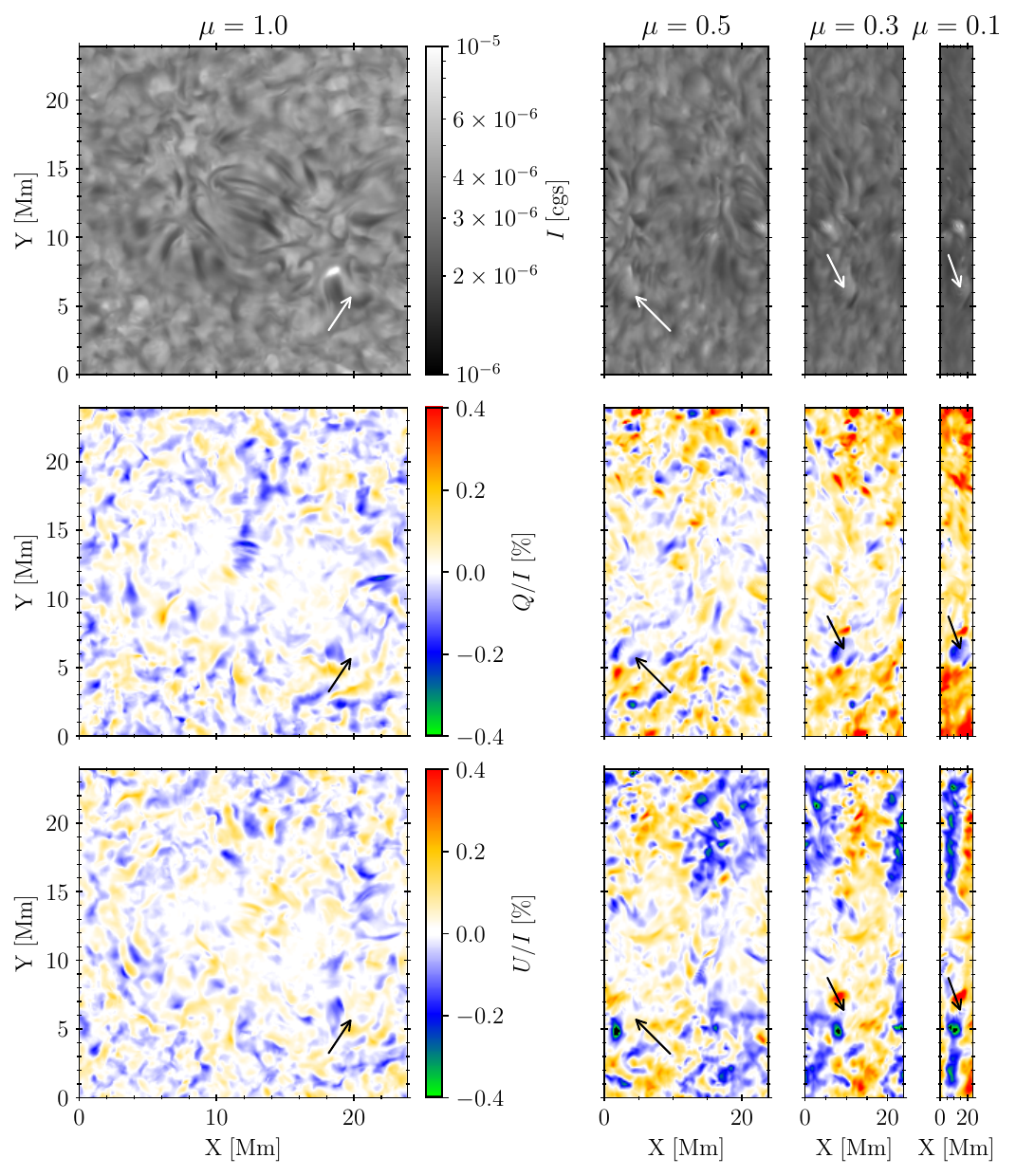}
\caption{
Maps of the Stokes $I$ (top row), $Q/I$ (middle row), and $U/I$ (bottom row) signals at the center of the $\ha$ line for four LOS inclinations: $\mu=1.0$, $\mu=0.5$, $\mu=0.3$, and $\mu=0.1$ (from left to right). The arrows in the Stokes $I$ maps highlight plasma structures associated with localized negative $Q/I$ signals, which are most prominent at $\mu=0.1$. The reference direction for positive Stokes $Q$ is parallel to the solar limb -- that is, along the $Y$ axis.
}
  \label{fig:stokes_maps}
\end{figure*}

Figure~\ref{fig:stokes_maps} shows the spatial distribution of the emergent Stokes signals at the line center for several LOS inclinations. At the disk center, both $Q/I$ and $U/I$ exhibit small-scale positive and negative patches produced by local symmetry-breaking effects. Toward the limb, the scattering geometry introduces a preferred direction, causing the $Q/I$ signals to become predominantly positive and increase in amplitude. In contrast, although the amplitudes of the $U/I$ signals also increase, their spatially averaged values remain close to zero because no preferred sign exists for the azimuthal symmetry breaking.

\begin{figure*}[h!]
  \centering
  \includegraphics[trim=20 50 0 20, clip,
    width=0.9\textwidth]{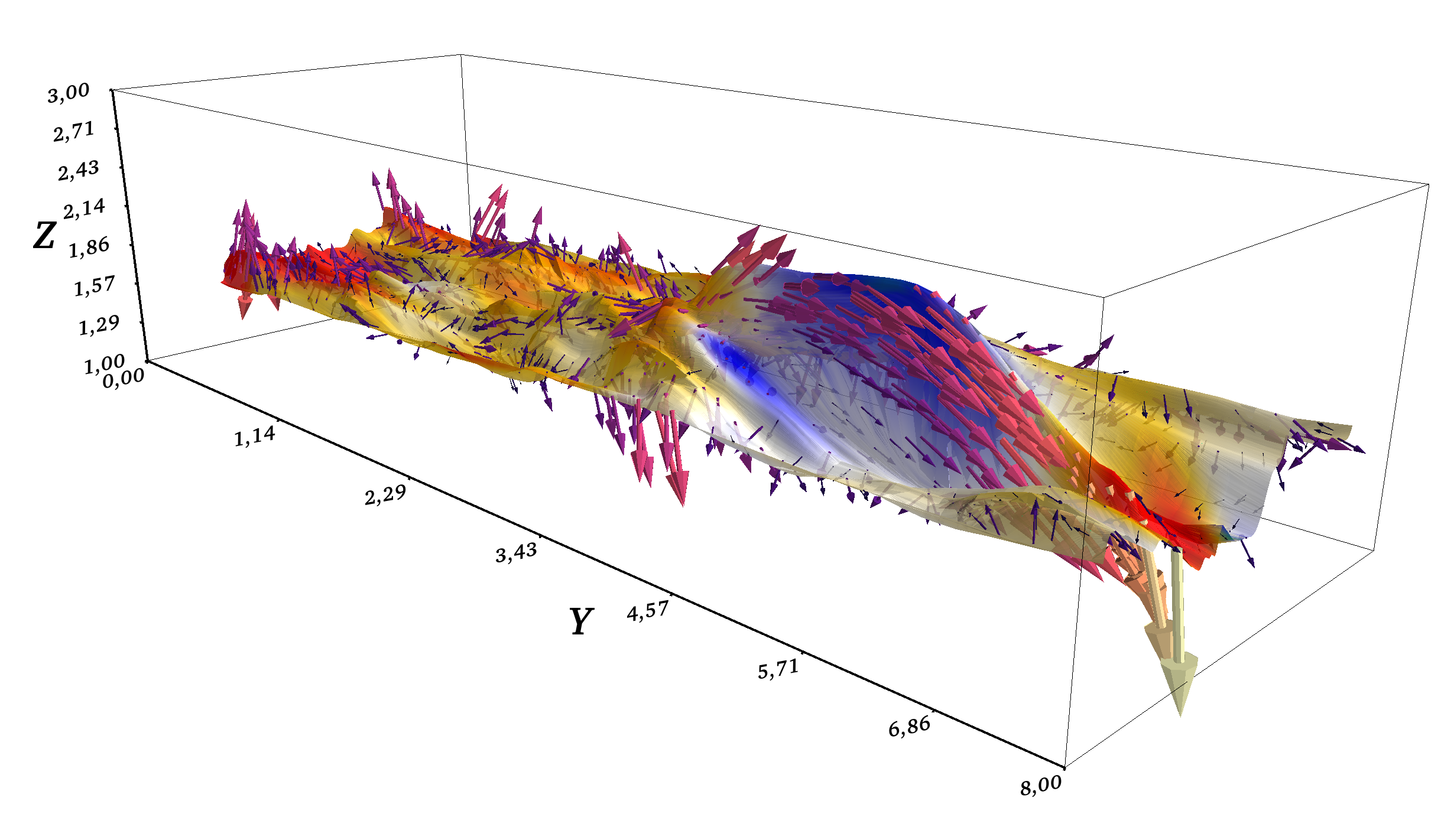}
  \vspace{-0.4cm}
  \includegraphics[trim=20 20 0 50, clip,
  width=0.9\textwidth]{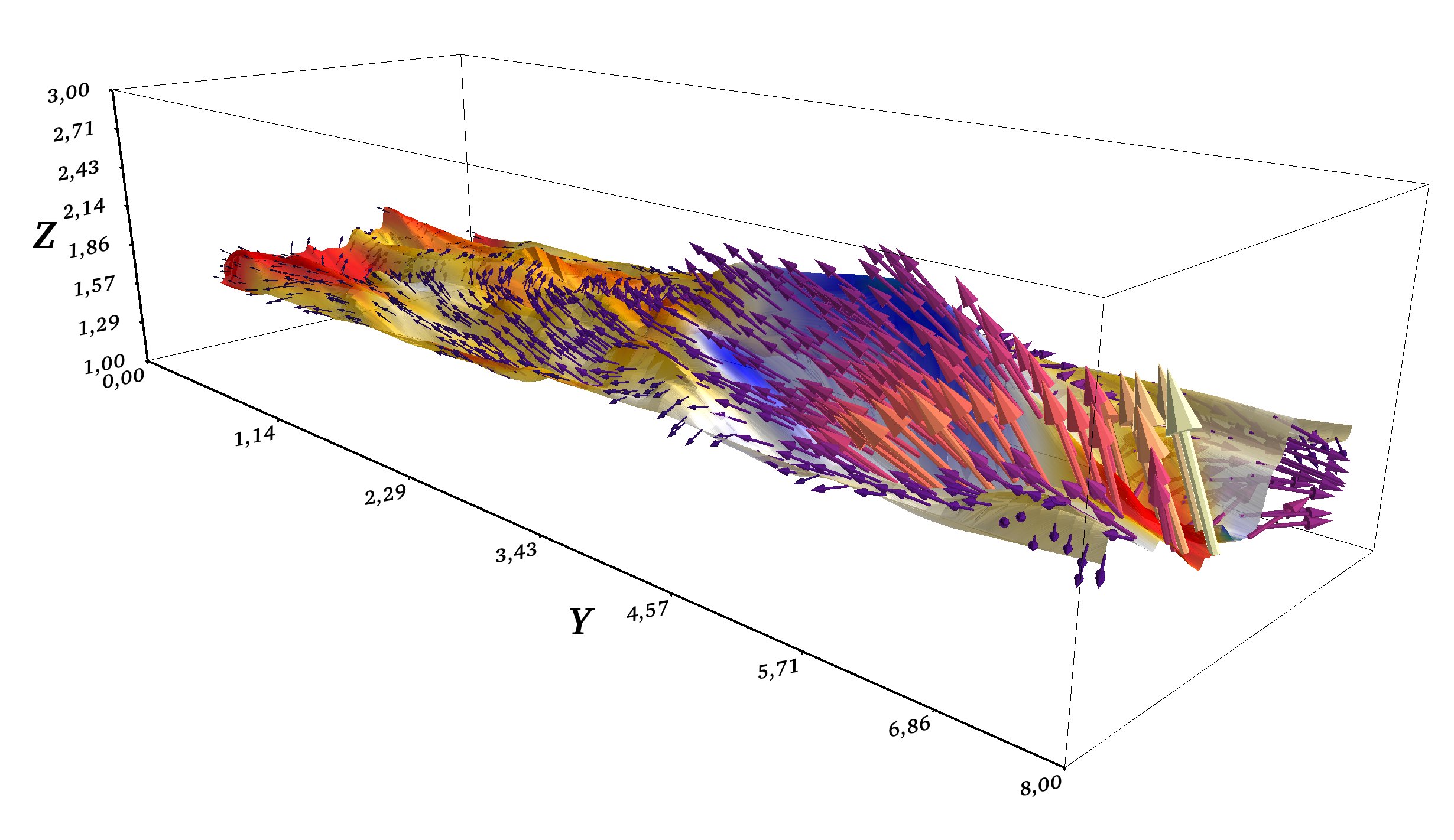}
  \caption{
Corrugated $\tau=1$ surface at the $\ha$ line center for $\mu=0.1$ in the 3D MHD model atmosphere. The surface is color-coded by the line-center $Q/I$ signal shown in Fig.~\ref{fig:stokes_maps}. The arrows represent the velocity field (top panel) and the magnetic field (bottom panel) evaluated at the local line-center formation height. Arrow brightness scales with field magnitude, with darker arrows indicating smaller values and brighter arrows larger ones.
}
  \label{fig:blue_bubble}
\end{figure*}

The spatially resolved calculations reveal a wide diversity of polarization signatures. Of particular interest is the localized region of negative $Q/I$ signals visible near $Y\approx6\,$Mm for $\mu=0.1$. As shown in Fig.~\ref{fig:blue_bubble}, this feature is associated with elevated line-formation heights ($\sim2.4\,$Mm), strong horizontal flows reaching approximately $20\,$km s$^{-1}$, and relatively organized magnetic structures. In contrast, the neighboring region around $Y\approx8\,$Mm, characterized by strong positive $Q/I$ signals, forms at lower heights ($\sim1.5\,$Mm) and is associated with downflows reaching $30\,$km s$^{-1}$. Similar density structures influencing the H$\alpha$ formation height have been reported by \citet{2012ApJ...749..136L} (see their Fig.~15).

These examples illustrate the highly corrugated nature of the H$\alpha$ formation surface and the complex relationship between the emergent polarization signals and the local atmospheric properties. Magnetic fields, velocity gradients, density inhomogeneities, and radiative-transfer effects interact nonlinearly, producing a rich variety of polarization signatures across the field of view. These results highlight the importance of accounting for the nonmagnetic causes of symmetry-breaking (i.e., horizontal inhomogeneities and macroscopic velocity gradients) when interpreting H$\alpha$ spectropolarimetric observations. For the Ca\,{\sc i} 4227~\AA\ line, \citet{2021ApJ...909..183J} show that an improved 1.5D approximation can reproduce qualitatively several properties obtained from full 3D calculations. Whether a similar approximation provides sufficient accuracy for H$\alpha$ has not been investigated here and remains to be established.

\section{Theory versus observations\label{sec:th_vs_obs}}

\begin{table}
  \centering
  \caption{
Observational parameters used to degrade the synthetic profiles to reproduce the instrumental and seeing conditions of the ZIMPOL observations.
}
  \label{tab:zimpol}
  \begin{tabular}{ll}
    \hline
    \hline
    Spatial sampling & $1.4''$ \\
    Spectral sampling & $9.7\,$m\r{A} \\
    Slit width & $0.5''$ \\
    Seeing & $1.5''$ \\
    Spatial pixels & 140\\
    Spectral pixels & 1240\\
    Spectral resolving power & $2.5\cdot10^{5}$ \\
    Telescope aperture & $0.45\,$m \\
    Polarimetric accuracy & $2\cdot10^{-4}$ \\
    \hline
  \end{tabular}
\end{table}

\begin{figure}[t]
  \centering
  \includegraphics[width=1\columnwidth]{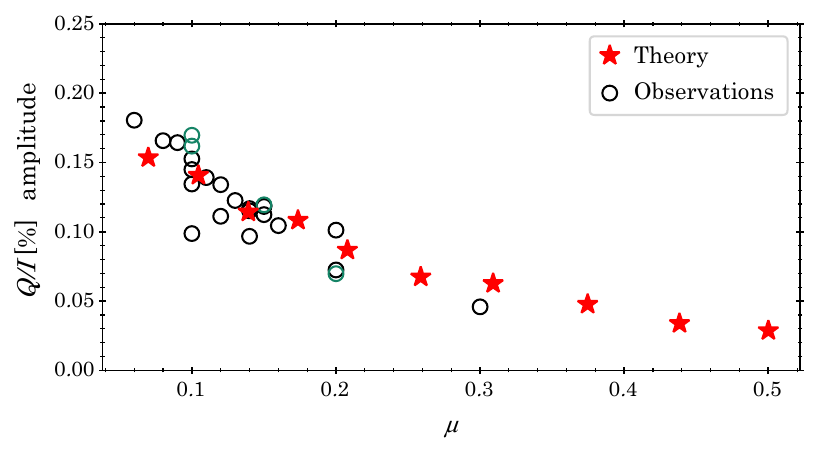}
\caption{
CLV of the $Q/I$ line-center amplitude at the north and south solar limbs. The open black circles and open green circles show the observations reported by \cite{2022A&A...659A.179J} at the north and south solar limbs, respectively, while the red symbols represent the synthetic profiles degraded according to the observational parameters listed in Table~\ref{tab:zimpol}. The reference direction for positive Stokes $Q$ is parallel to the solar limb.
}
  \label{fig:clv_th_obs}
\end{figure}

The Stokes profiles of the emergent radiation presented in Sect.~\ref{sec:fs} correspond to the intrinsic output of the 3D MHD simulation and therefore do not include the effects of finite spatial resolution, atmospheric seeing, instrumental broadening, or photon noise. To compare the theoretical predictions with spectropolarimetric observations, the synthetic profiles must be degraded to reproduce the characteristics of the observing system \citep[e.g.,][]{2018ApJ...863..164D,2021ApJ...909..183J,2021ApJ...909..180D}.

The H$\alpha$ observations used here were obtained with the ZIMPOL at the Gregory-Coudé Telescope of the IRSOL observatory in Locarno, Switzerland, and are described in \citet{2022A&A...659A.179J}. These observations achieve a polarimetric sensitivity on the order of $10^{-4}$ per pixel after approximately nine minutes of integration, with a spatial resolution of about $3''$ and a spectral resolution close to $20\,$m\AA. They reveal substantial spatial variability in the H$\alpha$ scattering polarization signals, providing a stringent observational constraint for assessing the realism of current 3D MHD models of the solar chromosphere.

Figure~\ref{fig:clv_th_obs} shows the observed center-to-limb variation (CLV) of the H$\alpha$ line-center $Q/I$ amplitude together with the degraded synthetic profiles. The theoretical calculations reproduce the observed overall center-to-limb trend and yield $Q/I$ amplitudes comparable to the measurements over the sampled range of viewing angles, indicating that the dominant physical mechanisms governing the formation of the H$\alpha$ scattering polarization are realistically captured by the model.

The discrepancies between the synthetic and observed $Q/I$ amplitudes may arise from several sources. The polarization amplitudes are sensitive to both Hanle depolarization and collisional depolarization, whose relative importance depends critically on the electron and proton densities. Since these densities are affected by the nonequilibrium ionization balance of hydrogen, uncertainties in the chromospheric ionization state may influence the predicted signals \citep{2016A&A...585A...4C}. Additional uncertainties are associated with the collisional rates between the fine-structure levels of H$\alpha$ (see Appendix~\ref{sb:ha_collisions}). Furthermore, the observations were obtained in coronal-hole regions, whereas the simulation represents an enhanced-network atmosphere. Differences in the magnetic topology sampled by the LOS are therefore expected. Additional observations at intermediate heliocentric angles would be valuable for assessing the robustness of the predicted CLV for $\mu>0.2$.

To simulate slit-based spectropolarimetric observations, a synthetic slit was projected onto the model atmosphere approximately parallel to the solar limb. To avoid artificial repetitions caused by the periodic horizontal boundaries of the computational domain, the slit was rotated by $5^\circ$ with respect to the $Y$-axis. The resulting synthetic spectrograms were then degraded according to the spatial, spectral, and polarimetric characteristics of the observational setup, following the procedure described by \citet{2021ApJ...909..183J} and using the parameters listed in Table~\ref{tab:zimpol} (see also \citealt{2004A&A...422..703G}; \citealt{10.1117/12.857120}).

\begin{figure*}[t]
  \centering
  \includegraphics[width=1\textwidth]{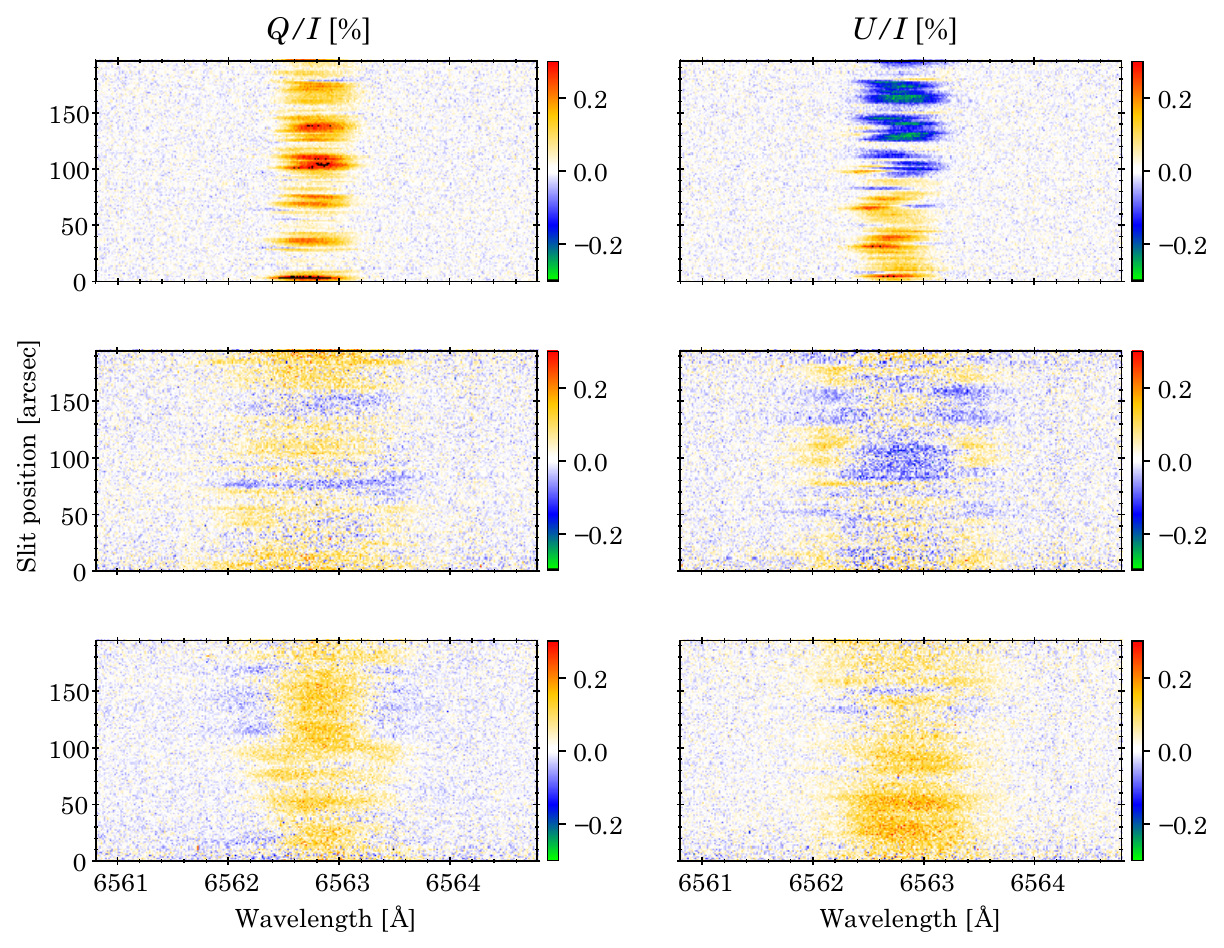}
\caption{
Comparison of the spatially resolved $Q/I$ and $U/I$ signals along the spectrograph slit at $\mu=0.14$. Top panels: Synthetic profiles degraded according to the observational parameters listed in Table~\ref{tab:zimpol}. Middle panels: Observations obtained at the east limb in a strong magnetic region. Bottom panels: Observations obtained at the west limb in a quiet region outside coronal holes. The exposure time for each observation is $\sim6$ minutes, yielding a polarimetric sensitivity of $\sim2\times10^{-4}$. The reference direction for positive Stokes $Q$ is parallel to the slit. The color scale saturates at $\pm0.3\,\%$.
}
  \label{fig:slit_degr}
\end{figure*}

Figure~\ref{fig:slit_degr} compares the degraded synthetic spectrograms with observations obtained in a magnetically active region and in a quieter region at the east and west limbs, respectively. Several similarities between theory and observations are immediately apparent. Both exhibit substantial spatial variability along the slit, with polarization amplitudes fluctuating on scales of only a few arc seconds. Moreover, regions of enhanced magnetic activity are systematically associated with reduced linear-polarization amplitudes, consistent with the action of Hanle depolarization. In the synthetic spectrograms, this spatial pattern repeats approximately every $30''$, reflecting the horizontal periodicity of the computational domain.

The main differences concern the amplitudes and widths of the polarization structures. The synthetic signals generally exhibit larger amplitudes and sharper spatial variations than the observations. These differences are expected because the calculations correspond to a single instantaneous snapshot, whereas the observations integrate over several minutes. Temporal averaging is known to reduce both the amplitudes and the spatial contrast of chromospheric scattering-polarization signals \citep{2016ApJ...831L...5C,2021ApJ...909..180D}.

The synthetic profiles also remain systematically narrower than the observed ones. This discrepancy most likely reflects the limited amount of unresolved small-scale dynamics present in the MHD simulation \citep{2016ApJ...826L..10S,2018A&A...619A..60J}. Stronger velocity gradients along the LOS would increase the effective Doppler broadening through differential Doppler shifts along the photon path, producing wider polarization signatures.

Overall, the comparison indicates that the present 3D calculations reproduce the overall observed center-to-limb trend and yield polarization amplitudes broadly consistent with the observations, while also capturing important aspects of the spatial organization of the H$\alpha$ scattering polarization. The remaining discrepancies may be associated with limitations of the atmospheric model and with the effects of the time evolution that cannot be captured by a single simulation snapshot.

\section{Asymmetries in the linear polarization profiles\label{sec:asym}}

\begin{figure}[t]
  \centering
  \includegraphics[width=0.45\textwidth]{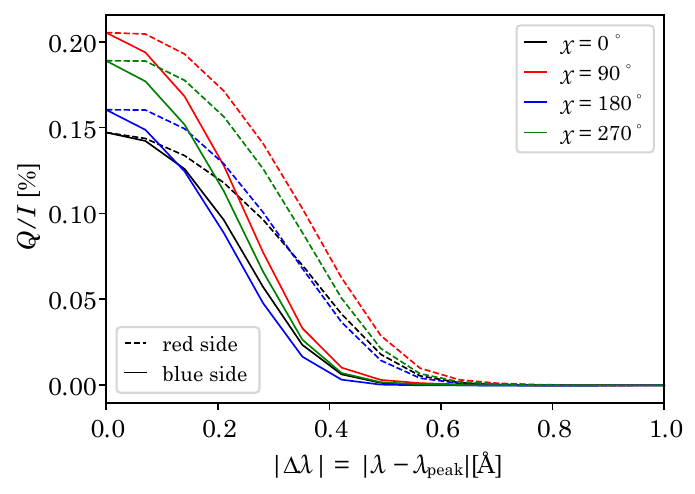}
\caption{
Comparison of the red- and blue-side Stokes $Q/I$ profiles at $\mu=0.1$ for different azimuth angles, $\chi$. The dashed and solid lines correspond to the red side ($\lambda-\lambda_{\mathrm{peak}}>0$) and blue side ($\lambda-\lambda_{\mathrm{peak}}<0$), respectively. The profiles are spatially averaged over all pixels for the selected LOS.
}
  \label{fig:qi_azimuth}
\end{figure}

The observed $Q/I$ profile at $\mu=0.1$ (Fig.~\ref{fig:avg_mus_profiles}) exhibits a clear asymmetry between the blue and red sides. A similar type of LCA
is found in the synthetic profiles computed for different azimuths $\chi$ at the same LOS. Figure~\ref{fig:qi_azimuth} illustrates this LCA by comparing representative $Q/I$ profiles for $\chi=0^\circ$, $90^\circ$, $180^\circ$, and $270^\circ$.
In this figure, the wavelength offsets are measured relative to the maximum of each polarization profile ($\Delta\lambda=\lambda-\lambda_{\mathrm{peak}}$), rather than to the nominal H$\alpha$ line center. With this choice, the polarization profiles are generally steeper in wavelength for $\Delta\lambda<0$ than for $\Delta\lambda>0$. The observed profile in the right panel of Fig.~\ref{fig:avg_mus_profiles} also exhibits qualitatively the same asymmetry.

The dependence of the emergent polarization profiles on the LOS azimuth is a direct consequence of the 3D nature of the model atmosphere. Horizontal inhomogeneities in temperature, density, velocity, and magnetic field break the axial symmetry of the radiation field, causing the optical path and the local anisotropy experienced by the scattering atoms to vary with azimuth. As a result, different viewing directions sample different atmospheric structures (see also Fig.~\ref{fig:blue_bubble}), giving rise to the variety of LCA patterns observed in the synthetic profiles.

The spatially resolved $Q/I$ signals across the field of view display a wide variety of shapes, ranging from nearly symmetric single-peaked profiles to strongly asymmetric and multi-peaked profiles. Since these profile morphologies encode information on the local atmospheric conditions, they may provide useful diagnostics of chromospheric magnetism and dynamics.

\begin{figure*}[t]
  \centering
  \includegraphics[width=0.95\textwidth]{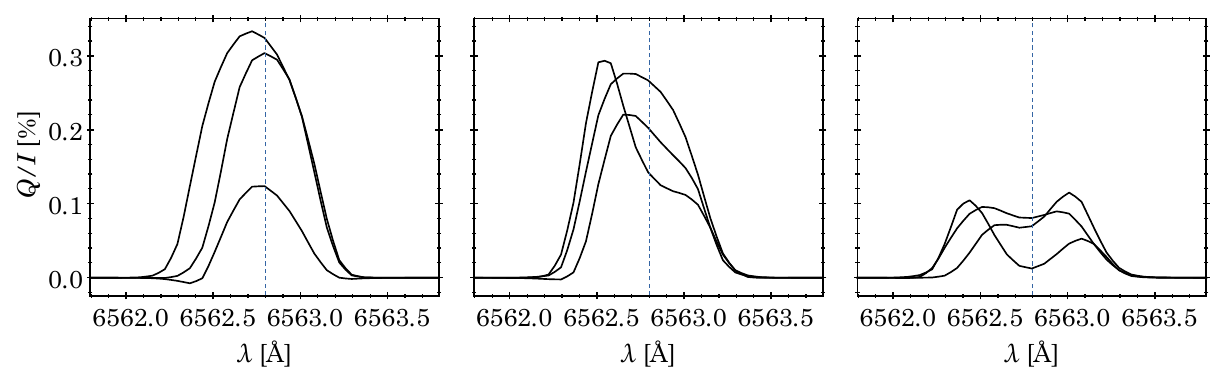}
\caption{
Representative synthetic Stokes $Q/I$ profiles at $\mu=0.1$, illustrating the adopted asymmetry classification. Left panel: Symmetric, single-peaked profiles (degree 0). Middle panel: Asymmetric, single-peaked profiles (degree 1). Right panel: Multi-peaked profiles (degree 2). The vertical dotted line indicates the reference line-center wavelength.
}
  \label{fig:degrees_asym}
\end{figure*}

To quantify these variations, we classify the synthetic $Q/I$ profiles into three categories, illustrated in Fig.~\ref{fig:degrees_asym}: (0) symmetric single-peaked profiles, (1) asymmetric single-peaked profiles, and (2) multi-peaked profiles. The classification procedure consists of:

\begin{enumerate}

\item interpolating each synthetic profile onto a uniformly spaced wavelength grid;

\item normalizing each profile by its maximum positive $Q/I$ amplitude, so that its peak amplitude is unity;

\item identifying all positive peaks with amplitudes larger than $2\%$ of the profile maximum;

\item excluding from the classification profiles containing negative peaks, which exceed $10\%$ of the maximum positive amplitude;

\item for single-peaked profiles:

\begin{enumerate}

\item computing the absolute difference between the areas under the peak-normalized $Q/I$ profile on either side of its maximum;

\item assigning degree 0 when the area difference falls below a lower threshold;

\item assigning degree 1 when the area difference exceeds an upper threshold;

\item leaving profiles with intermediate values unclassified;

\end{enumerate}

\item assigning degree 2 to profiles containing multiple positive peaks.

\end{enumerate}

Profiles containing substantial negative $Q/I$ signals are excluded from this particular morphological classification because the criteria defining degrees 0--2 are designed to characterize the symmetry and peak structure of profiles dominated by positive line-core polarization. Such negative profiles are physically meaningful, as illustrated by the localized signals discussed in connection with Figs.~\ref{fig:stokes_maps} and \ref{fig:blue_bubble}, but constitute a different morphological population that is not considered in the present classification.

Unsupervised clustering techniques, such as $k$-means clustering, provide an alternative approach for classifying spectropolarimetric profiles
\citep[e.g.,][]{2025ApJ...987L..39M}. Here, we adopt a deliberately simple morphology-based classification because our aim is to isolate a small number of readily interpretable profile properties---symmetry and number of peaks---and relate them to the atmospheric quantities in their formation regions.

\begin{figure}[t]
  \centering
  \includegraphics[width=1\columnwidth]{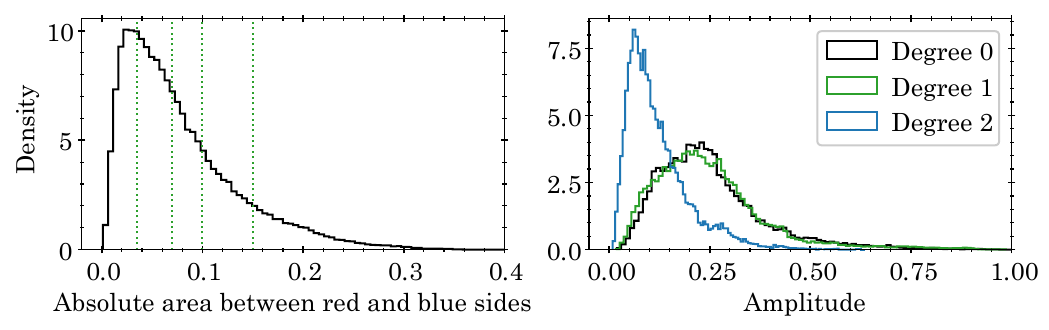}
\caption{
Left panel: Histogram of the absolute area difference between the red and blue sides of single-peaked Stokes $Q/I$ profiles. The gray-shaded region corresponds to degree-0 profiles, while the dotted green lines indicate the thresholds adopted to define the degree-1 profiles. Right panel: Distribution of the profile amplitudes for each asymmetry class, using an area-difference threshold of 0.1 to define the degree-1 profiles. The amplitudes shown in the right panel correspond to the maximum positive $Q/I$ values of the individual profiles, expressed in percent, before normalization for the asymmetry classification.
}
  \label{fig:diff_ampl}
\end{figure}

The adopted thresholds suppress spurious peaks and exclude profiles dominated by negative signals, providing a practical separation between the different profile classes. Figure~\ref{fig:diff_ampl} summarizes the statistical properties of the classification. Approximately $16.6\%$ of the profiles belong to degree 0, whereas the fraction of degree-1 profiles varies between $7.9\%$ and $50\%$ depending on the adopted asymmetry threshold. About $24.5\%$ of the profiles are classified as degree 2.

The amplitude distributions reveal a clear distinction between the different profile classes. Symmetric and asymmetric single-peaked profiles exhibit similar amplitudes, both peaking around $0.25\%$. In contrast, multi-peaked profiles show substantially weaker signals, with the distribution peaking near $0.05\%$. This difference is consistent with stronger Hanle depolarization in the regions where degree-2 profiles form, although other atmospheric effects may also contribute to their morphology and amplitudes.

\begin{figure}[t]
  \centering
  \includegraphics[width=1\columnwidth]{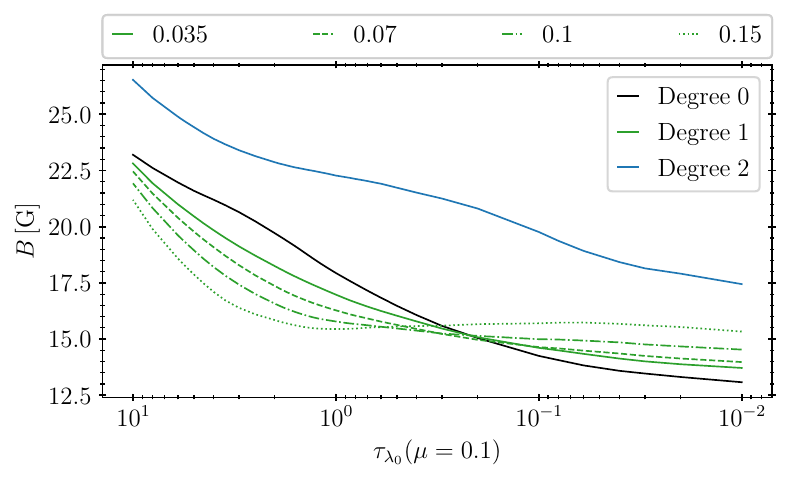}
\caption{
Variation of the mean magnetic field strength with optical depth, $\tau_{\mu=0.1}(\lambda_0)$, for profiles classified according to their degree of asymmetry. The green lines show single-peaked asymmetric profiles (degree 1) obtained using different classification thresholds. The black line corresponds to symmetric single-peaked profiles (degree 0). The blue lines show multi-peaked profiles (degree 2), which exhibit stronger magnetic fields throughout the line-formation region.
}
  \label{fig:bv_tau_asym}
\end{figure}

To investigate the possible role of Hanle depolarization, we examined the magnetic-field strength as a function of line-center optical depth for each profile class. Figure~\ref{fig:bv_tau_asym} shows the corresponding mean magnetic-field stratifications for different asymmetry thresholds.

Symmetric profiles (degree 0) exhibit a gradual decrease in the magnetic-field strength with height, reaching approximately $18\,$G at $\tau=1$. Asymmetric single-peaked profiles (degree 1) show a somewhat flatter stratification, with typical values around $16\,$G in the line-formation region. In contrast, multi-peaked profiles (degree 2) are systematically associated with stronger magnetic fields, with average strengths of approximately $22.5\,$G at $\tau=1$.

The differences in amplitude and magnetic-field strength among the three profile classes can be interpreted, at least in part, in terms of the Hanle effect. The critical Hanle fields of the upper levels involved in the strongest H$\alpha$ transitions are $16.2\,$G, $9.2\,$G, and $6.1\,$G for the $3p_{3/2}$, $3d_{3/2}$, and $3d_{5/2}$ levels, respectively \citep{2004ASSL..307.....L}. The magnetic fields associated with degree-2 profiles therefore exceed the Hanle critical fields of the strongest H$\alpha$ transitions, consistent with significant Hanle depolarization, which may contribute to the occurrence of multi-peaked profiles. Degree-1 profiles are associated with magnetic-field strengths for which differential Hanle depolarization among the overlapping fine-structure components can contribute to the asymmetry between the blue and red sides.

Degree-0 profiles should not be interpreted as corresponding to an absence of the Hanle effect. Their mean magnetic-field strength at $\tau=1$, approximately $18\,$G, is comparable to or larger than the critical Hanle fields of the upper levels of the strongest H$\alpha$ components. Their nearly symmetric shape may therefore result from the combined action of the magnetic field, velocity gradients, and 3D symmetry-breaking effects, whose contributions may partially compensate in the emergent profile. The present results establish statistical correlations between profile morphology and the magnetic properties of the line-formation region, but they do not imply that the magnetic field alone determines the three profile classes. A clean separation of the magnetic contribution from those of macroscopic velocities and horizontal atmospheric inhomogeneities would require additional calculations in which the magnetic field is artificially set to zero while keeping the thermodynamic and velocity structure unchanged. Such an experiment is beyond the scope of the present work.

\section{Conclusions}

We presented a comprehensive 3D non-LTE radiative transfer investigation of the intensity and scattering polarization of the solar H$\alpha$ line in a realistic MHD model of the solar atmosphere. The calculations were carried out with the PORTA code using a multilevel hydrogen atom that accounts for the atomic polarization of the levels involved in the Ly$\alpha$, Ly$\beta$, and H$\alpha$ transitions. Our calculations provide a realistic 3D forward-modeling framework for investigating the formation of the H$\alpha$ scattering polarization in the inhomogeneous and dynamic solar chromosphere.

The synthesized intensity profiles are narrower and deeper than the observed ones, indicating that the 3D MHD model still lacks some of the small-scale dynamics and thermodynamic properties of the real solar chromosphere. In spite of these limitations, the calculated linear polarization signals capture several key observational properties of the H$\alpha$ line. In particular, the theoretical CLV reproduces the observed overall trend and yields line-center $Q/I$ amplitudes broadly consistent with the available observations, supporting the ability of the adopted atmospheric model and radiative transfer treatment to capture the main properties of the observed scattering polarization.

The spatially resolved calculations reveal a rich variety of polarization signals produced by the combined action of the Hanle effect, macroscopic plasma motions, and symmetry-breaking effects caused by the 3D atmospheric structure. Near the solar limb, the $Q/I$ amplitudes reach values on the order of $0.5\%$, while localized regions with negative $Q/I$ signals are associated with strong flows and enhanced line-formation heights. These results show that the scattering polarization of H$\alpha$ is highly sensitive to the local chromospheric conditions.

Another key result concerns the LCAs of the synthetic $Q/I$ profiles and their relationship with the atmospheric properties of their formation regions. By classifying the synthetic profiles according to their morphology and relating them to the magnetic properties of their formation regions, we find statistical correlations between profile morphology and magnetic-field strength that are consistent with an important role of differential Hanle depolarization among the individual H$\alpha$ fine-structure components. Multi-peaked profiles preferentially originate in regions with relatively strong magnetic fields, whereas single-peaked asymmetric profiles are associated with weaker magnetic fields with a comparatively flatter stratification throughout the line-formation region.
Symmetric profiles, on the other hand, are comparatively rare and are typically associated with regions where the magnetic-field strength decreases gradually with height. Their occurrence at magnetic-field strengths comparable to the critical Hanle fields 
shows that a nearly symmetric profile does not imply the absence of the Hanle effect. More generally, the present calculations do not allow the effects of magnetic fields to be completely disentangled from those of velocity gradients and horizontal atmospheric inhomogeneities.

These results highlight the importance of the nonmagnetic causes of symmetry breaking in the formation and interpretation of the H$\alpha$ scattering polarization. Horizontal inhomogeneities, macroscopic velocity gradients, and magnetic fields interact nonlinearly, producing polarization signatures that are captured by the present 3D radiative transfer calculations. Whether simplified approaches, such as the improved 1.5D approximation shown to reproduce qualitatively several properties of the scattering polarization of the Ca\,{\sc i} 4227~\AA\ line, can reproduce the essential properties of the H$\alpha$ polarization remains to be investigated. Although direct inference of the chromospheric magnetic field from H$\alpha$ observations remains challenging, the present study provides further insight into polarized line formation in the solar atmosphere.

The H$\alpha$ line provides a valuable diagnostic for future 3D spectropolarimetric investigations of chromospheric magnetism and dynamics. Its sensitivity to magnetic fields over an extended range of chromospheric heights, together with its response to collisional and dynamical effects, provides valuable constraints for 3D non-LTE inversions \citep[e.g.,][]{2022A&A...659A.137S,2025A&A...699A..73S,2025ApJ...987L..39M} and for the next generation of high-sensitivity solar observations.

\begin{acknowledgements}
This research was carried out within the framework of the European Research Council (ERC) Advanced Grant Agreement No.~742265 (POLMAG; https://research.iac.es/proyecto/polmag/). J.J.B. and J.T.B. acknowledge financial support from the Agencia Estatal de Investigación (AEI), Ministerio de Ciencia, Innovación y Universidades (MCIU), through grant PID2022-136563NB-I00, co-funded by the European Regional Development Fund (ERDF). The Instituto de Astrofísica de Canarias has been accredited as Centro de Excelencia Severo Ochoa, a distinction awarded by the Spanish State Research Agency for the period 2026–2031. J.\v{S}. acknowledges financial support from project RVO:67985815 of the Astronomical Institute of the Czech Academy of Sciences. The 3D radiative transfer calculations were carried out on the MareNostrum supercomputer at the Barcelona Supercomputing Center (BSC-CNS, Spain). We gratefully acknowledge the technical support provided by the Spanish Supercomputing Network (RES). Additional computational resources were provided by the La Palma Supercomputer at the Instituto de Astrofísica de Canarias.
\end{acknowledgements}

\bibliographystyle{aa}
\bibliography{ms.bib}

\begin{appendix}
\nolinenumbers

\section{Statistical equilibrium and the $\rhokq$ elements}
\label{sb:see_halpha}

In quiet regions of the solar atmosphere, the linear polarization of the H$\alpha$ line is dominated by scattering processes and the ensuing atomic alignment. Although the Zeeman effect may generate measurable circular polarization, its contribution to the linear polarization is generally negligible. We therefore neglect atomic orientation and retain only the multipolar components associated with atomic alignment.

Under these conditions, the levels with $J=1/2$ are described solely by the population component $\rho^0_0$, while the levels with $J=3/2$ require the six components $\rho^0_0$, $\rho^2_0$, $\rho^2_{\pm1}$, and $\rho^2_{\pm2}$. In principle, the level $3d_{5/2}$ also admits multipolar components with rank $K=4$, increasing the total number of independent density-matrix elements to 38.

However, the solar radiation field is only weakly anisotropic, namely
\begin{equation}
\frac{|J^K_Q(\nu)|}{J^0_0(\nu)} \ll 1,
\qquad K=1,2.
\end{equation}
As discussed by \citet{2004ASSL..307.....L} and \citet{2011ApJ...732...80S}, this implies that the multipolar components with $K>2$ are negligibly small compared with those of rank $K=2$. Consequently, the $K=4$ components of the $3d_{5/2}$ level can be neglected, yielding an accurate description of the atomic excitation with only 29 independent $\rho^K_Q$ elements.

This reduction substantially simplifies the statistical-equilibrium problem while preserving the accuracy required for modeling the scattering polarization of the H$\alpha$ line in weakly magnetized regions of the solar chromosphere.

\section{Radiative transfer and the $\jkq$ components}
\label{sb:rte_halpha}

Following \citet{2011ApJ...732...80S}, the radiative transfer problem is formulated in the Hanle regime, neglecting the Zeeman splittings in the statistical-equilibrium equations. Under this approximation, transitions involving only levels with $J=1/2$ contribute exclusively to the isotropic component of the radiation field, $J^0_0$, because they cannot generate or absorb linear polarization. The remaining transitions contribute to the full set of radiation field tensors required to describe scattering polarization. As a result, 46 independent $J^K_Q$ components are required to characterize the radiation field interacting with the atomic system.

The continuum opacity and emissivity are also included in the radiative transfer calculations. Since their frequency variation is small across the spectral regions considered here, these quantities are evaluated only at the central frequencies of the Ly$\alpha$, Ly$\beta$, and H$\alpha$ transitions. The corresponding continuum coefficients were computed with the RH code \cite[][]{2001ApJ...557..389U} using a hydrogen atom with three principal levels plus continuum and neglecting fine-structure effects.

This treatment provides the continuum contribution required for the non-LTE solution while keeping the computational complexity of the 3D calculations at a manageable level.

\section{Collisions}
\label{sb:ha_collisions}

Collisions play a crucial role in the non-LTE formation of the polarized H$\alpha$ line because they affect both the level populations and the atomic polarization. The collisional treatment adopted in this work closely follows that described by \citet{2011ApJ...732...80S}, and only a brief summary is given here.

We account for electron-impact excitations and de-excitations between hydrogen levels with different principal quantum numbers,
$\dtrans{H(n)+e}{H(n')+e'}$.
The corresponding rates are computed using the electron collisional strengths calculated by \citet{2004ApJ...609.1181P}. Following \citet{2011ApJ...732...80S}, these rates are converted into collisional probabilities between the fine-structure levels of the atomic model and expressed in the irreducible tensor formalism of the atomic density matrix \cite[][]{2004ASSL..307.....L}.

We also include collisional transitions between fine-structure levels belonging to the same principal quantum number. Although the associated energy separations are very small, these processes efficiently redistribute population among the fine-structure sublevels and produce a significant depolarization of the atomic alignment. As discussed by \citet{1986A&A...156...79B} and \citet{2011ApJ...732...80S}, the dominant transitions satisfy the selection rule $\Delta l=\pm1$.

The corresponding collisional rates are computed using the semi-classical theory of \citet{1996A&A...309..317S}, following the implementation described by \citet{2011ApJ...732...80S}. Both population-transfer ($K=0$) and alignment-transfer ($K=2$) rates are included in the statistical-equilibrium equations. The explicit expressions for these rates, as well as their formulation in terms of irreducible tensor components, can be found in Appendix~4 of \citet{2004ASSL..307.....L} and in \citet{2011ApJ...732...80S}.

\section{Boundary conditions and initialization}
\label{sb:boundary_halpha}

At the upper boundary of the computational domain, no incoming radiation is assumed. At the lower boundary, the incident radiation is specified by the Planck function at the local temperature. This approximation is justified because the lower boundary is placed sufficiently deep in the atmosphere for the Ly$\alpha$, Ly$\beta$, and H$\alpha$ lines to be effectively thermalized. Under these conditions, the source function approaches the Planck function and the atomic populations remain close to their thermodynamic-equilibrium values. Periodic boundary conditions are imposed in both horizontal directions ($x$ and $y$), consistently with the periodic horizontal boundaries of the 3D MHD model. Radiation leaving the computational domain through one horizontal boundary therefore re-enters through the corresponding opposite boundary.

The initialization of the atomic density matrix is important for the convergence of the non-LTE iterative scheme. Rather than adopting LTE populations as the initial guess, we initialize the statistical-equilibrium equations from the collisional solution obtained in the absence of radiation, assuming that all atoms are initially in the ground level. Preliminary numerical experimentation showed that this choice improves the convergence properties of the iterative solution for the H$\alpha$ problem compared with a standard LTE initialization.

\section{The preconditioning technique}
\label{sb:precond_halpha}

The statistical-equilibrium and radiative-transfer equations were solved iteratively using the Jacobi scheme implemented in PORTA \cite[][]{2013A&A...557A.143S}. For the present hydrogen model, the procedure follows the formalism for multilevel atoms with overlapping transitions explained in \citet{2011ApJ...732...80S}.

Because the H$\alpha$ line is formed by several blended fine-structure transitions, the standard Jacobi iteration must be modified to account for the coupling between overlapping line components. This is achieved through a preconditioning strategy in which the radiation field tensors are expressed as linear corrections to their values from the previous iteration. The resulting coupling coefficients depend on the overlap of the line profiles and on the approximate $\Lambda$ operator.

Substituting these preconditioned radiation field tensors into the statistical-equilibrium equations introduces nonlinear terms involving products of density-matrix elements. Following the approach explained in \citet{2011ApJ...732...80S}, these terms are linearized by replacing one of the factors by its value from the previous iteration. This procedure preserves the linearity of the statistical-equilibrium problem while retaining the accelerated convergence properties of the Jacobi scheme.

Numerical experimentation showed that the inclusion of the overlap preconditioning is important for achieving efficient and stable convergence of the H$\alpha$ polarized radiative-transfer problem.

\section{Implementation in the PORTA code}
\label{sb:implementation_halpha}

The H$\alpha$ radiative transfer problem was solved with the PORTA code \cite[][]{2013A&A...557A.143S}, using a dedicated atomic module developed for the hydrogen model described in this paper. The module computes the radiative rates, the radiative transfer coefficients, the radiation field tensors $\bar{J}^{K}_{Q}$, and the solution of the statistical-equilibrium equations for the atomic density matrix. Macroscopic plasma velocities are fully taken into account through the corresponding Doppler shifts in the observer's frame.

The original 3D MHD model was reduced before performing the radiative transfer calculations in order to make the computations tractable while retaining the atmospheric regions relevant for the formation of the Ly$\alpha$, Ly$\beta$, and H$\alpha$ lines. The horizontal resolution was reduced by a factor of two, resulting in a grid of $252\times252$ points. In addition, the deepest layers of the model, where the hydrogen lines are fully thermalized, and the highest coronal layers, where they become effectively transparent, were removed. The resulting computational domain contains $252\times252\times337$ grid points and spans approximately $7.1\,$Mm in height, extending from $0.6\,$Mm below the visible solar surface to $6.5\,$Mm in the low corona.\footnote{The height $z=0$ corresponds to the photospheric layer where $\tau_{5000}=1$.}

The angular integration was performed using the quadrature scheme implemented in PORTA, consisting of a Gaussian quadrature in the inclination angle and a trapezoidal quadrature in azimuth. The adopted discretization comprises 128 propagation directions, corresponding to four inclinations and four azimuths per octant.

The frequency grids were optimized to accurately sample both the line cores and the extended wings of the hydrogen lines. Uniform frequency spacing was used in the line cores, while a logarithmic spacing was adopted in the wings. The final grids contain 61 frequency points for the Ly$\alpha$ and Ly$\beta$ lines and 99 frequency points for H$\alpha$.

The adopted spatial, angular, and spectral discretizations were selected after extensive numerical experimentation to balance accuracy and computational cost. The resulting calculations were carried out on the MareNostrum and La Palma supercomputers using the domain-decomposition strategy of PORTA.

\end{appendix}

\end{document}